\documentclass[9pt,twocolumn,twoside]{optica}
\setboolean{shortarticle}{false}
\setboolean{minireview}{false}
\usepackage{bm}
\usepackage{comment}
\usepackage{braket}
\newcommand\inv[1]{#1\raisebox{1.15ex}{$\scriptscriptstyle-\!1$}}

\title{A Photonic Crystal Slab Laplace Differentiator}

\author[1]{Cheng Guo}
\author[2]{Meng Xiao}
\author[2]{Momchil Minkov}
\author[2]{Yu Shi}
\author[2,*]{Shanhui Fan}

\affil[1]{Department of Applied Physics, Stanford University, Stanford, California 94305, USA}
\affil[2]{Ginzton Laboratory and Department of Electrical Engineering, Stanford University, Stanford, California 94305, USA}
\affil[*]{Corresponding author: shanhui@stanford.edu}

\dates{Compiled October 21, 2017}

\ociscodes{(050.5298) Photonic crystals; (100.1160) Analog optical image processing.}

\doi{\url{http://dx.doi.org/XXXX/XXXX}}

\begin{abstract}
We introduce an implementation of a Laplace differentiator based on a photonic crystal slab that operates at transmission mode. We show that the Laplace differentiator can be implemented provided that the guided resonances near the $\Gamma$ point exhibit an isotropic band structure.  Such a device may facilitate nanophotonics-based optical analog computing for image processing.  
\end{abstract}

\setboolean{displaycopyright}{false}

\begin{document}

\maketitle

\section{Introduction}
\label{sec:intro}
In image processing, spatial differentiation accomplishes image sharpening and edge-based segmentation, with broad applications ranging from microscopy and medical imaging to industrial inspection and object detection.\cite{Gonzales2008, Markham1963, Abramoff2004, Brosnan2004,Dalal2005} In these applications, of particular importance are isotropic derivative operators, whose response is rotationally invariant. The simplest and most widely used isotropic derivative operator is the two-dimensional Laplacian.\cite{Rosenfeld1982}

Spatial differentiation can certainly be carried out with conventional digital electronic computation. However, there are many big-data applications that require real-time and high-throughput image processing, for which digital computations become challenging.\cite{Pham2000,Holyer1989} Optical analog computing may overcome this challenge by offering high-throughput derivative operation with almost no energy consumption other than the inherent optical loss associated with the differential operators.\cite{Solli2015,Gorlitz1975,Eu1973,Reinhardt1978,Sirohi1977,Kasprzak1982,Yao1971,Nesteruk2001,Warde1983,Davis1991,Lancis1997} Moreover, the recent developments of nanophotonic structures such as meta-surfaces and plasmonic structures have offered the possibility of doing optical differentiation using compact devices. 
However, most of the existing works on optical spatial differentiation using nanophotonic structures are restricted to one dimension, whereas  most images are two-dimensional objects.\cite{Zhu2017, Golovastikov2015, Golovastikov2014, Silva2014, Youssefi2016, AbdollahRamezani2015, Pors2015} To date, there has been no optical realization of the two-dimensional Laplacian at transmission mode with a compact nanophotonic device. 
\begin{figure}
\centering
\includegraphics[width=0.9\columnwidth]{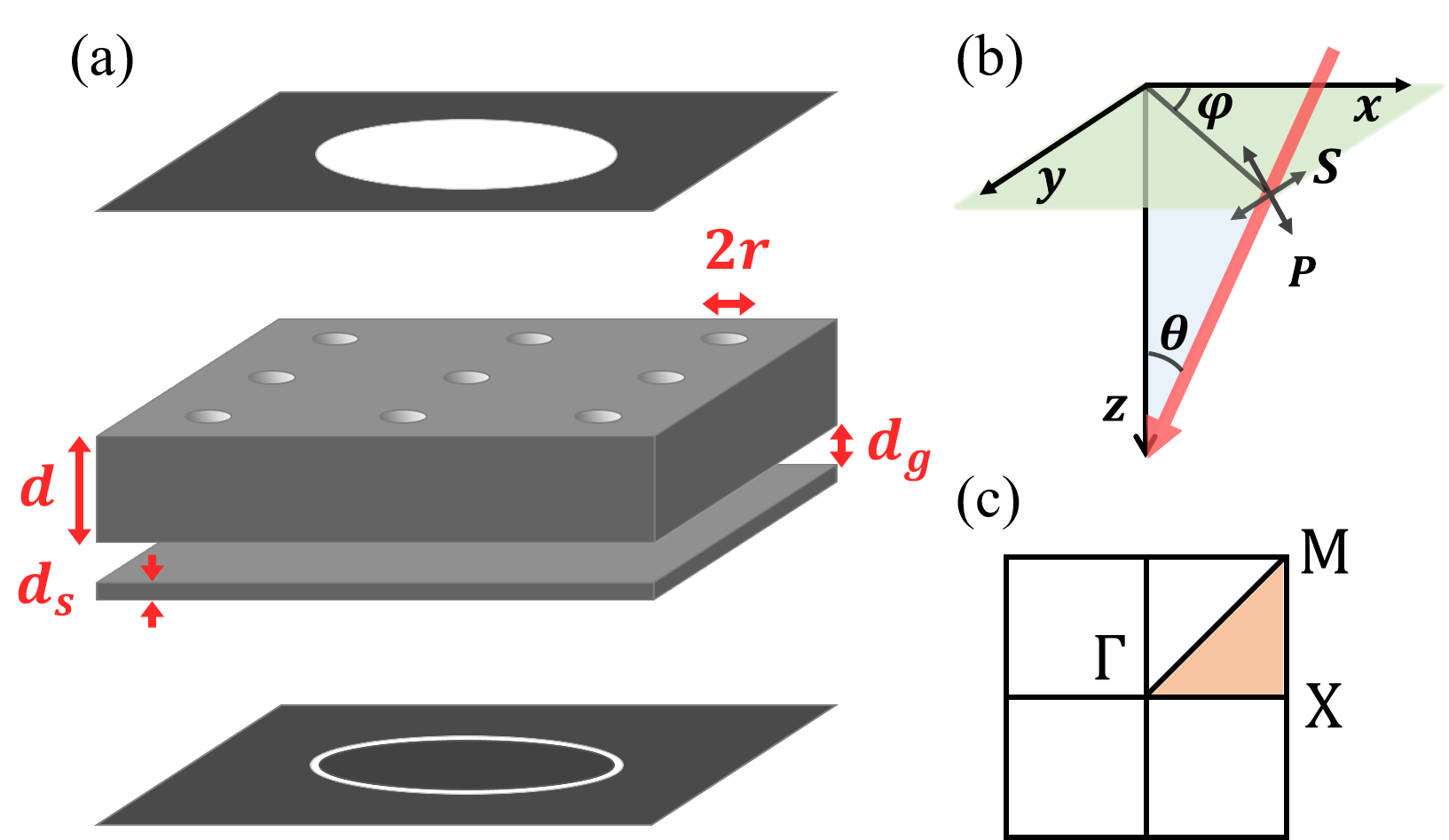}
\caption{\label{fig1}(a) Geometry of the photonic crystal slab differentiator, which consists of a photonic crystal slab separated from a uniform dielectric slab by an air gap. For $\epsilon=12$, the geometry parameters are: $d=0.55a$, $r=0.111a$, $d_s=0.07a$, $d_g=0.21a$. The plane above and below shows the input and output image respectively. The transmitted image (e.g. ring) through the device is the Laplacian of the incident image (e.g. disk) illuminated by a normally incident light with frequency $\omega_0 = 0.47656 \times 2\pi c/a$. (b) Coordinate system. (c) The Brillioun zone of the system.}
\end{figure} 

In this article, we show that a two-dimensional Lapacian can be implemented with the use of a photonic crystal slab, one of the most widely studied nanophotonic structures.\cite{Chow2000,Joannopoulos2011,Zhou2014} As shown in Figure~\ref{fig1}, the structure consists of a photonic crystal slab separated from a uniform dielectric slab by an air gap. The key in this design is to achieve an unusual band structure that is isotropic for the two polarizations.
\section{Theoretical Analysis}
\label{sec:analysis}
Our objective is to realize a Lapacian for a two-dimensional input field. Specifically, for a normally incident light beam along the $z$-axis with a transverse field profile $S_{in}(x,y)$, we would like to design an optical device for which the transmitted beam has a profile $S_{out}(x,y) \propto \nabla^2 S_{in}(x,y)$, whre $\nabla^2 = \partial_x^2 + \partial_y^2$ is the Laplacian. The task of realizing the Laplacian $\nabla^2$ in the real space is equivalent to designing an optical system with response function
\begin{equation}
t(\bm{k}) \propto (k_x^2+k_y^2) \label{eq:t-propto-k}
\end{equation} in the wavevector space ($\bm{k}$-space).\cite{Bracewell1986} 
Equation~(\ref{eq:t-propto-k}) requires that $t(\bm{k})=0$ at $|\bm{k}|=0$. In a lossless photonic crystal slab, the transmission for normally incident light can vanish near a guided resonance.\cite{Fan2002} Thus we consider in more details the transmission coefficient of a photonic crystal slab near normal incidence. Consider a single photonic band of guided resonances, as characterized by $\bm{k}$-dependent resonant frequencies $\omega(\bm{k})$ and radiative linewidths $\gamma(\bm{k})$. (Here $\bm{k} = (k_x, k_y)$ refers to the in-plane wavevector.) Near the resonant frequencies,
 the transmitted amplitude $t$ is expressed as \cite{Fan2002}
\begin{equation}
t(\omega,\bm{k}) = t_d + f \frac{\gamma(\bm{k})}{i[\omega-\omega(\bm{k})]+\gamma(\bm{k})}~, \label{eq:t_origin}
\end{equation}
where $\omega$ is the incident light frequency, $t_d$ is the direct transmission coefficient, and $f$ is related to the complex decaying amplitude of the resonance to the transmission side of the slab.

In general, $f$ is constrained by the direct process due to energy conservation and time-reversal symmetry.\cite{Fan2002,Wang2013} In particular, if $t_d =1$, that is, the direct 
pathway has a $100\%$ transmission coefficient, then 
\begin{equation}
f = -t_d = -1 \label{eq:f_td}
\end{equation}
even for structure without z-mirror symmetry such as shown in Figure \ref{fig1}, as has been derived in Ref. \cite{Wang2013}. In this special case, 
\begin{equation}
t(\omega,\bm{k}) = 1 - \frac{\gamma(\bm{k})}{i[\omega-\omega(\bm{k})]+\gamma(\bm{k})}~. \label{eq:t_special}
\end{equation}
We denote $\omega_0 = \omega(\bm{k}=\bm{0})$, $\gamma_0 = \gamma(\bm{k}=\bm{0})$. Using Equation~(\ref{eq:t_special}), zero transmission occurs at the $\Gamma$ point when the incident wave has the frequency:
\begin{equation}
	\omega = \omega_0 ~.
\end{equation}
At $\omega = \omega_0$, we perform an expansion of the transmission coefficient $t$ near $\bm{k}=\bm{0}$ :
\begin{equation}
	t(\omega_0, \bm{k}) = 0 + \frac{\partial t}{\partial \omega(\bm{k})}\Bigr|_{\Gamma} \delta \omega(\bm{k}) + \frac{\partial t}{\partial \gamma(\bm{k})}\Bigr|_{\Gamma} \delta \gamma(\bm{k}) ~, \label{eq:t}    
\end{equation}
where
\begin{equation}
\delta \omega(\bm{k}) = \omega(\bm{k}) - \omega_0,\qquad \delta \gamma(\bm{k}) = \gamma(\bm{k})-\gamma_0 ~,
\end{equation}
\begin{equation}
\frac{\partial t}{\partial \omega(\bm{k})}\Bigr|_{\Gamma} = -\frac{i}{\gamma_0},\qquad \frac{\partial t}{\partial \gamma(\bm{k})}\Bigr|_{\Gamma} = 0~,
\end{equation}
therefore,
\begin{equation}
t(\omega_0, \bm{k}) = - \frac{i}{\gamma_0}\delta \omega(\bm{k})~. 
\end{equation}
In this special case, $t(\bm{k})$ is simply proportional to the band dispersion $\delta\omega(\bm{k})$ near $\bm{k}=\bm{0}$. If $\delta\omega(\bm{k}) \propto |\bm{k}|^2$, then $t(\bm{k})\propto |\bm{k}|^2$ as well. Therefore the transmission through a photonic crystal slab can be used to achieve the Laplacian.
\begin{figure*}
\centering
\includegraphics[width=1.5\columnwidth]{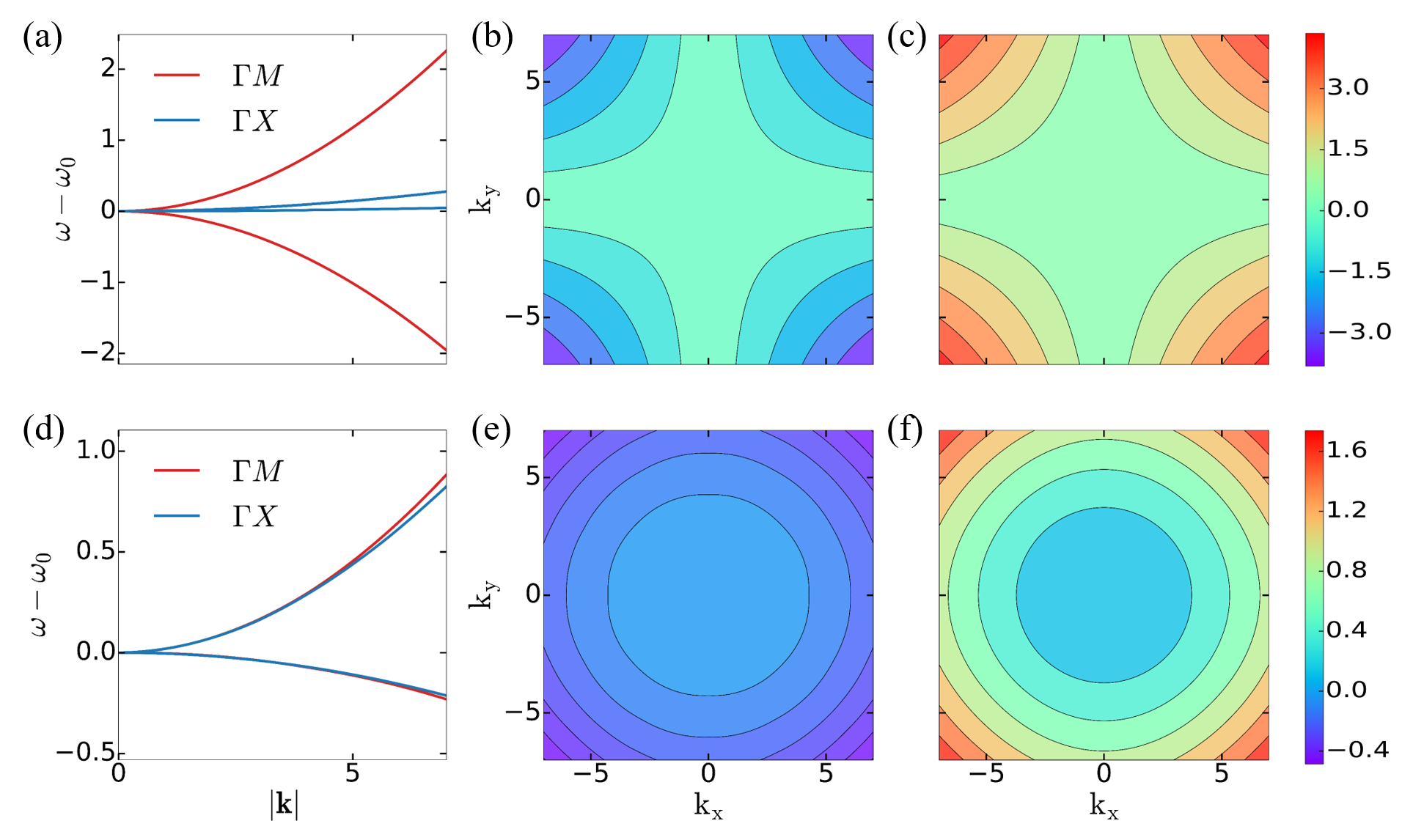}
\caption{\label{fig2} Band structure of the photonic crystal slab with the dielectric constant  $\epsilon=12$, the thickness $d=0.55a$, and the radius $r=0.111a$ of the holes. (a-c) Band structure near $\omega_0= 0.38749 \times 2\pi c/a$ as a typical example of conventional anisotropic bands that are doubly degenerate at $\Gamma$. (a) Band dispersion  diagram along $\Gamma\textrm{-}X$ and $\Gamma\textrm{-}M$. (b) Constant frequency contours of the lower band with respect to $(k_x,k_y)$. (c) Constant frequency contours of the upper band. (d-f) Band structure at $\omega_0= 0.47656 \times 2\pi c/a$, which is nearly isotropic. (d) Band dispersion along $\Gamma\textrm{-}X$ and $\Gamma\textrm{-}M$. (e) Constant frequency contours of the lower band. (f) Constant frequency contours of the upper band. The frequency $(\omega-\omega_0)$ is in the units of $10^{-4}\times2\pi c/a$, while $|\bm{k}|$, $k_x$ and $k_y$ are in the units of $10^{-3}\times2\pi/a$.}
\end{figure*}

The analysis above indicates that, to use a photonic crystal slab as a two-dimensional spatial differentiator, it is sufficient that the slab satisfies the following three conditions:  
\begin{enumerate}
\item $t_d = 1$.
\item Only one guided resonance band is coupled.
\item The band satisfies the dispersion $(\omega(\bm{k})-\omega_0) \propto |\bm{k}|^2$.
\end{enumerate}

To satisfy the first condition above, we note that the direct transmission coefficient $t_d$ is related to the non-resonant transmission pathway \cite{Fan2002}. Hence it is possible to realize $t_d = 1$ by, e.g. changing the thickness of the slab. In the structure as shown in Figure~\ref{fig1}, we achieve $t_d = 1$ by placing a uniform dielectric slab in the vicinity of the photon crystal slab, and by tuning the distance between the slabs. This has the advantage that we can tune $t_d$ without significantly affecting the band structure of the photonic crystal slab.

To satisfy the second and third conditions above, one will need to design the band structure for the photonic crystal slab. The design here is in fact quite non-trivial due to the vectorial nature of  electromagnetic waves. Since the Laplacian is isotropic in $\bm{k}$-space, it is natural to consider a photonic crystal slab structure that has rotational symmetry. As an illustration, here we consider a slab structure with a square lattice of air holes that has $C_{4v}$ symmetry. For such a slab, it is known that  at the $\Gamma$ point, which corresponds to $|\bm{k}|=0$, the only modes that can couple to external plane wave must be two-fold degenerate, belonging to a two-dimensional irreducible representation of the $C_{4v}$ group.\cite{Ochiai2001,Fan2002} Near such  modes, in the vicinity of $\Gamma$ point, the band structure in general can be described by the following $2 \times 2$ effective Hamiltonian: (See Supplementary Materials)
\begin{equation} \label{eq:H-k}
\hat{\mathcal{H}}(\bm{k}) = (\omega_0 - i \gamma_0 + a |\bm{k}|^2)\hat{\mathbf{I}} + b (k_x^2-k_y^2) \hat{\bm{\sigma}}_z+ c k_x k_y \hat{\bm{\sigma}}_x~,
\end{equation}
where $a,b,c$ are three complex coefficients, the $\bm{\sigma}$'s are the Pauli matrices. This Hamiltonian has two eigenvalues of
\begin{equation} 
\omega_{\pm}(\bm{k}) - i \gamma_{\pm}(\bm{k}) = \omega_0 - i \gamma_0 + a|\bm{k}|^2 \pm \sqrt{b^2(k_x^2-k_y^2)^2+ c^2 k_x^2k_y^2} ~.\label{eq:eigenvalue}
\end{equation}

The band structure as described by Equation~(\ref{eq:eigenvalue}) in general does not satisfy the conditions for ideal spatial differentiation as outlined above. In this paper for concreteness we will fix the dielectric constant of the material for the slabs to be  $\epsilon = 12$, which approximates that of Si or GaAs in the infrared wavelength range. In Figure~\ref{fig2}(a-c), we plot the band structure for a  slab with a thickness $d=0.55a$, and a radius  $r=0.111a$  of the holes, in the vicinity of a guided resonance at the $\Gamma$ point with a frequency $\omega_0= 0.38749 \times 2\pi c/a$. The band structure is numerically determined using guided-mode expansion method \cite{Andreani2006,Minkov2014}, and it agrees excellently with the analytic expression of Equation~(\ref{eq:eigenvalue}) (See Supplementary Materials).  At the $\Gamma$ point, the bands are two-fold degenerate. Away from the $\Gamma$ point, the degeneracy is lifted, and the two bands are strongly anisotropic, as can be seen in Figure~\ref{fig2}(a), where the effective masses along the $\Gamma \textrm{-}X$ and the $\Gamma \textrm{-}M$ direction are drastically different. The band anisotropy can also be visualized in the constant frequency contour plotted in Figure~\ref{fig2}(b) and (c), for the two bands. In general, it can be shown that, for a general choice of parameters $a$, $b$ and $c$, the constant frequency contour as described by Equation (\ref{eq:eigenvalue}) is not a circle even at the $|\bm{k}|\to0$ limit. 

On the other hand, Equation~(\ref{eq:eigenvalue}) also indicates that when $c=\pm2b$, both bands will be isotropic:
\begin{equation} 
\omega_{\pm}(\bm{k}) - i \gamma_{\pm}(\bm{k}) = \omega_0 - i \gamma_0 + (a\pm b)|\bm{k}|^2 ~.\label{eq:eigenvalue_iso}
\end{equation}
To achieve such an isotropic band structure requires detailed tuning of the parameters. Numerically, we found that such an isotropic band structure can be approximately achieved with the same slab as indicated above, but near the guided resonance at $\Gamma$ with a different frequency $\omega_0 = 0.47656 \times 2\pi c/a$, where the complex coefficients $a=0.68-0.14i,b=1.11-0.12i,c=2.24+0.15i$,  thus $c/2b=0.99 + 0.17i\approx1$. (See Supplementary Materials for details.) The resulting isotropic band structures are shown in Figure~\ref{fig2}(d-f). In Figure~\ref{fig2}(d), we see that the bands along the $\Gamma \textrm{-}M$ and $\Gamma \textrm{-}X$ direction have almost identical effective masses. And in Figure~\ref{fig2}(e) and (f), we see that the constant frequency contours are almost completely circular.    	

As we mentioned above, for a structure with $C_{4v}$ symmetry, the modes that a normally-incident plane wave can couple to at the $\Gamma$ point are always two-fold degenerate. Consequently, near the $\Gamma$ point, there are always two bands of guided resonance present. Moreover, off the normal direction, if the direction of incident waves is away from the high symmetry planes, both $S$ and $P$ polarized lights may couple to both bands, leading to complex polarization conversion effects. Thus,  in general, in addition to having an anisotropic band structure near $\Gamma$, a photonic crystal slab structure also does not satisfy the condition 2 above regarding the excitation of a single guided resonance band. Remarkably, however, below we show that once the condition for an isotropic band is satisfied, each of the two bands in fact only  couple to one single polarization, \textit{for every direction of incidence}. Below, we refer to this effect, where each polarization excites only a single band away from normal incidence, as the effect of \textit{single-band excitation}. 

To illustrate the effect of single-band excitation for every direction when the bands are isotropic, we first show that single-band excitation always occurs when the incident direction is in a mirror symmetry plane of the structure. Using the group theory notation in Ref.~\citep{Sakoda2005}, as our system possesses $C_{4v}$ symmetry, the doubly degenerate states at $\Gamma$ are $E$ modes. Along the $\Gamma\textrm{-}X$ direction, this pair of $E$ modes splits into two singly degenerate states of $A$ and $B$ modes which are even and odd, respectively, with respect to the reflection operator associated with the mirror plane $y=0$. On the other hand, since the $S$ and $P$ polarized lights are also odd and even with respect to the same mirror plane, respectively, along the $\Gamma\textrm{-}X$ direction, the $S$($P$) polarized light can only couple to the $B$($A$) modes. Thus, in general, we have the effect of single-band excitation when the direction of incidence is in a high symmetry plane such as the $y=0$ plane.\cite{Tikhodeev2002}

Next, we prove the following statement: if the two-band Hamiltonian is isotropic, that is,
\begin{equation}\label{eq:hamilton}
 \quad \hat{\mathcal{R}}(\varphi) \hat{\mathcal{H}}(\bm{k}) \hat{\mathcal{R}}^{-1}(\varphi) = \hat{\mathcal{H}}(\bm{R}(\varphi)\bm{k})~,
\end{equation}
for every $\varphi \in (0, 2\pi)$, then we have the effect of single-band excitation along all directions. Here, $\bm{R}(\varphi)$ and $\hat{\mathcal{R}}(\varphi)$ are the rotation operators that describe the rotation around the $z$-axis by an angle $\varphi$, in the $\bm{k}$ space and the two-dimensional Hilbert space, respectively. To prove this, we denote the eigenstates of the two bands as $\ket{\bm{k}, A}$ and $\ket{\bm{k}, B}$, which connect to the $A$ and $B$ modes along the $\Gamma \textrm{-}X$ direction respectively. Equation~(\ref{eq:hamilton}) implies that:
\begin{equation}\label{eq:eigenmodes} \begin{split}
	\ket{\bm{R}(\varphi)\bm{k}, A} &= \hat{\mathcal{R}}(\varphi) \ket{\bm{k}, A}~, \\
    \ket{\bm{R}(\varphi)\bm{k}, B} &= \hat{\mathcal{R}}(\varphi) \ket{\bm{k}, B}~. 
\end{split}
\end{equation}
We denote the $S$ and $P$ polarized modes as $\ket{\bm{k}, S}$ and $\ket{\bm{k}, P}$ respectively. By definition,
\begin{equation}\label{eq:SP} \begin{split}
	\ket{\bm{R}(\varphi)\bm{k}, S} &= \hat{\mathcal{R}}(\varphi) \ket{\bm{k}, S}~, \\
    \ket{\bm{R}(\varphi)\bm{k}, P} &= \hat{\mathcal{R}}(\varphi) \ket{\bm{k}, P} ~.
\end{split}
\end{equation}
For $\bm{k}=\bm{k_x}$ along the $\Gamma \textrm{-}X$ direction, $A$($B$) mode does not couple to the $S$($P$) polarizations, i.e.  
\begin{equation} \begin{split}
\braket{\bm{k_x}, S|\bm{k_x}, A}&=0~, \\
\braket{\bm{k_x}, P|\bm{k_x}, B}&=0~. 
\end{split}
\end{equation}
Then for $\bm{k} = \bm{R}(\varphi)\bm{k_x}$ along any direction, using Equations (\ref{eq:eigenmodes}) and (\ref{eq:SP}), we also have
\begin{equation} \begin{split}
\braket{\bm{k}, S|\bm{k},A}= \bra{\bm{k_x}, S}\inv{\hat{\mathcal{R}}}(\varphi)\hat{\mathcal{R}}(\varphi)\ket{\bm{k_x}, A} = 0 ~, \\
\braket{\bm{k}, P|\bm{k},B}= \bra{\bm{k_x}, P}\inv{\hat{\mathcal{R}}}(\varphi)\hat{\mathcal{R}}(\varphi)\ket{\bm{k_x}, B} = 0 ~.
\end{split}
\end{equation}

\begin{figure}
\centering
\includegraphics[width=1.0\columnwidth]{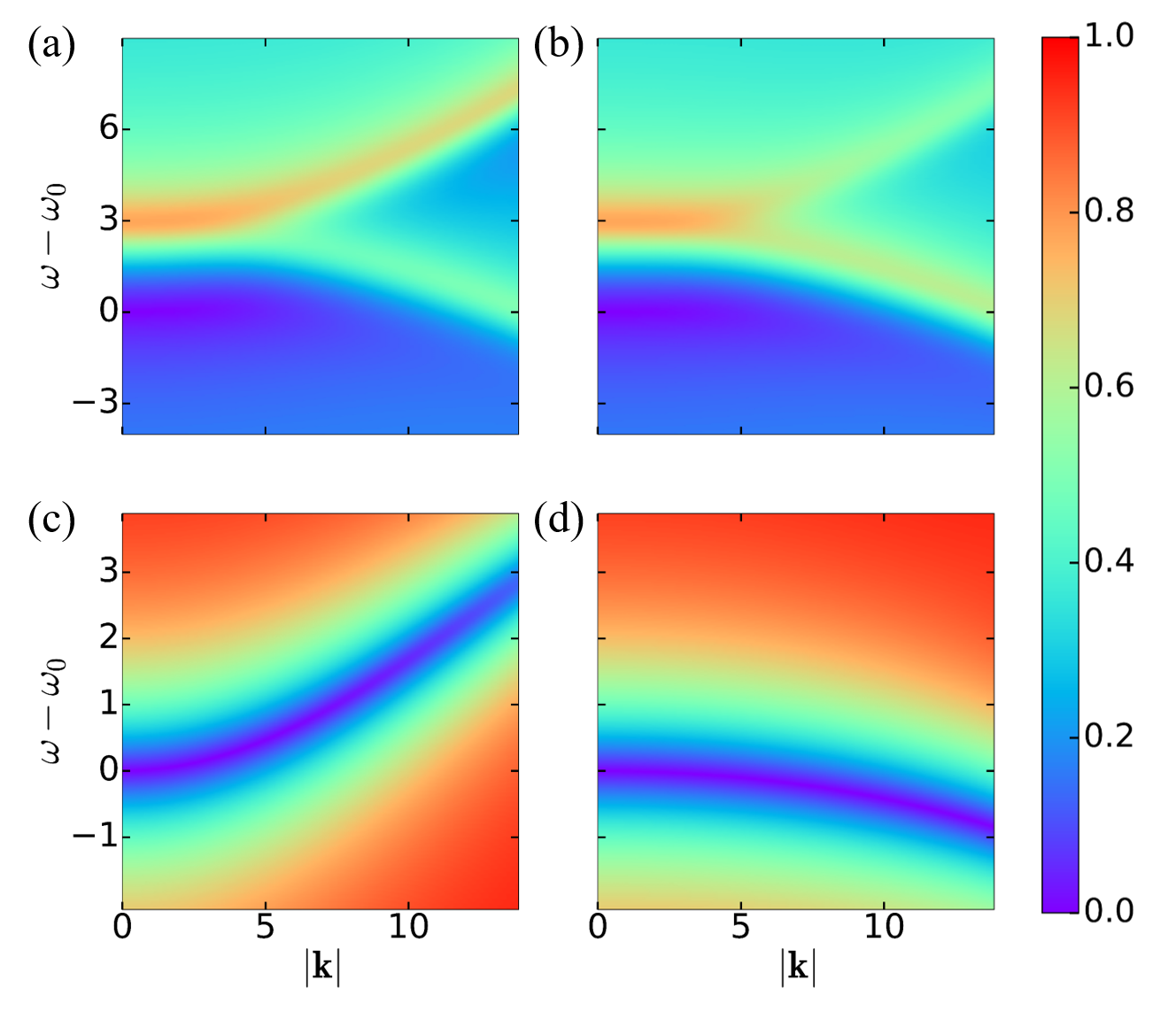}
\caption{\label{fig3}Transmittances of the two-slab structure as illustrated in Figure~\ref{fig1}. The photonic crystal slab has the same structure parameters as those of Figure~\ref{fig2}. The thickness of the uniform dielectric slab $d_s=0.07a$. The air gap width $d_g=2.1a$. (a,b) Transmittance near $\omega_0= 0.38749 \times 2\pi c/a$ as a function of $|\bm{k}|$ and frequency $(\omega-\omega_0)$ along a general direction $\varphi=14^\circ$ for (a) $S$ light and (b) $P$ light. Both bands couple to $S$ and $P$ light. (c,d) Transmittance near $\omega_0= 0.47656 \times 2\pi c/a$ along $\varphi=14^\circ$ for (c) $S$ light and (d) $P$ light. $S$ light only couples with the upper $B$ band while $P$ with the lower $A$ band, showing the single-band excitation effect. Frequency $(\omega-\omega_0)$ is in the units of $10^{-4}\times2\pi c/a$, while $|\bm{k}|$ is in the units of $10^{-3}\times2\pi/a$.}
\end{figure}
As a numerical demonstration of the connection of the single-band excitation effect with the isotropic band structure, we consider the two-slab structure as illustrated in Figure~\ref{fig1}. The photonic crystal slab structure has the same parameters as those of Figure~\ref{fig2}. The uniform dielectric slab has a thickness of $d_s=0.07a$. The air gap between the two slabs has a width of $d_g=2.1a$. Figure~\ref{fig3} shows the transmission coefficients through the two slab system, as we vary the in-plane wavevector, while maintaining $\varphi = 14^\circ$, and hence the direction of the incident light is away from any symmetry plane. Near  the frequency of $\omega = 0.38749 \times 2\pi c/a$, which is near the guided resonances as shown in Figure~\ref{fig2}(a-c), the band structure is strongly anisotropic, and both the $S$ and $P$ polarizations excite both bands as shown in Figure~\ref{fig3}(a-b). On the other hand, near the frequency of $\omega = 0.47656 \times 2\pi c/a$, which is near the guided resonances as shown in Figure~\ref{fig2}(d-f), each polarization excites only one band as shown in Figure~\ref{fig3}(c-d). 

In this structure, the thickness of the uniform dielectric slab and the gap size are chosen such that $t_d = 1$ near $\omega = 0.47656 \times 2\pi c/a$. Therefore, we have designed a structure that satisfies all the sufficient conditions for achieving an ideal Laplacian spatial differentiator. 

\begin{figure}
\centering
\includegraphics[width=1.0\columnwidth]{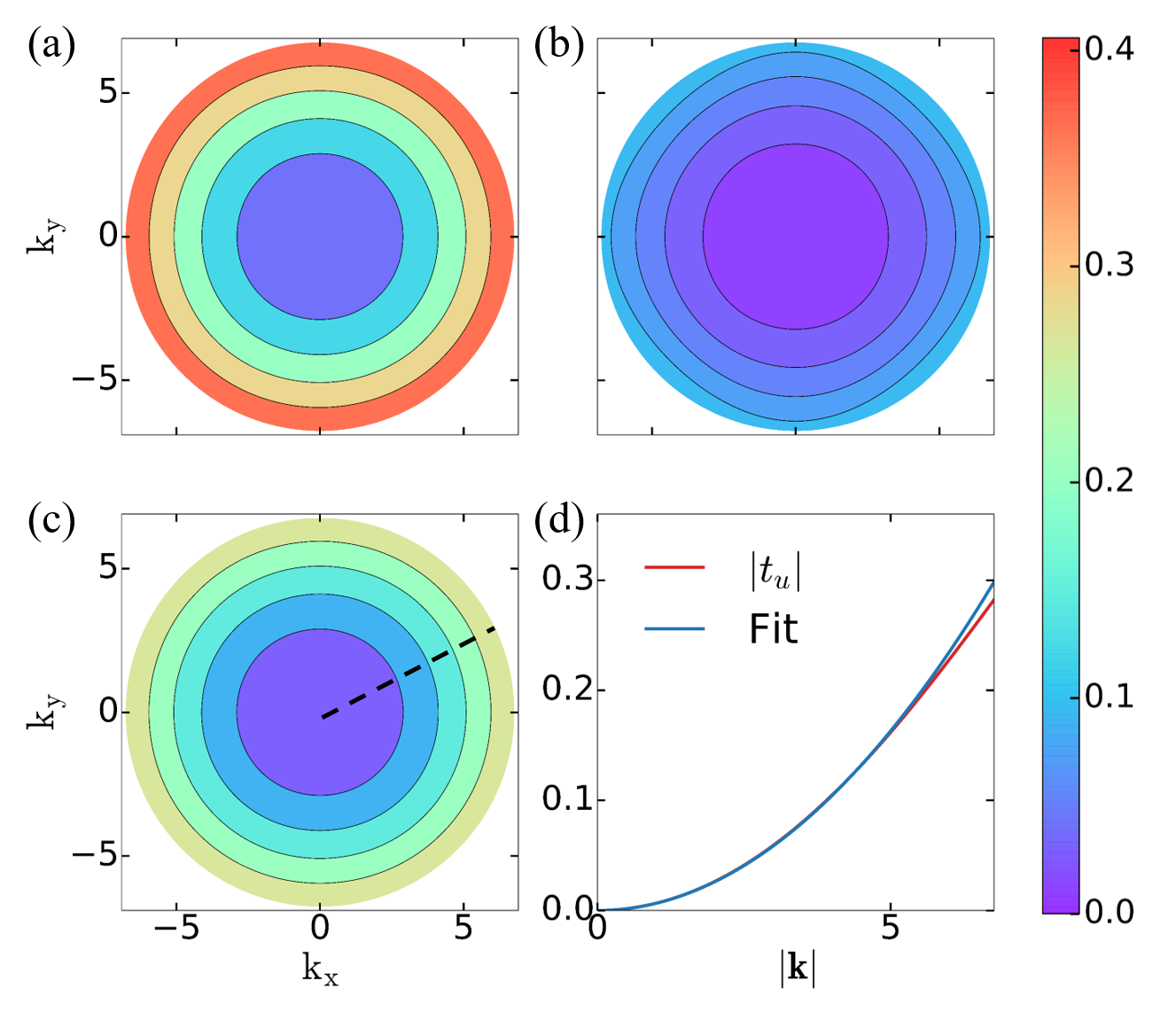}
\caption{\label{fig4}(a,b,c) Contour plot of transmittance $|t|$ as a function of $k_x$ and $k_y$ at frequency $\omega_0 = 0.47656 \times 2\pi c/a$ for (a) $S$, (b) $P$ and (c) unpolarized light, which are all isotropic near $\Gamma$. (d) $|t_u|$ as a function of $|\bm{k}|$ along $\varphi=14^\circ$ direction, and the quadratic fitting $t_u = \alpha |\bm{k}|^2$ where $\alpha$ is the only fitting parameter. $k_x$ and $k_y$ are in the units of $10^{-3}\times2\pi/a$.}
\end{figure}

In Figure~\ref{fig4}(a) and (b), we plot the transmission coefficients, as a function of in-plane wavevector $\bm{k}$, for the $S$ and $P$ polarized incident light, at the frequency $\omega = 0.47656 \times 2\pi c/a$, for the same structure used in Figure~\ref{fig3}. Both polarizations exhibit a transmission coefficient that is proportional to $|\bm{k}|^2$. In Figure~\ref{fig4}(c), we plot the transmission coefficient for unpolarized light at the same frequency. The transmission coefficient for the unpolarized light can be derived as (See Supplementary Materials for a derivation)
\begin{equation}
|t_u| = \sqrt{\frac{|t_s|^2+|t_p|^2}{2}}~. 
\end{equation}
Here $t_s$ and $t_p$ are the transmission coefficients for $S$ and $P$ light respectively. The transmission coefficient for the unpolarized light also shows an isotropic response. In Figure~\ref{fig4}(d), we plot $|t_u|$ as a function of $|\bm{k}|$,  which shows a quadratic dependency. To show this, we fit the curve with quadratic function $|t_u|=\alpha|\bm{k}|^2$, where $\alpha$ is the only fitting parameter. The fitting is almost perfect for $|\bm{k}|$ up to $6\times10^{-3}\times2\pi/a$, which as we will show provides a sufficient range of wavevector for image differentiation. Thus, the transmission coefficients have all the required $\bm{k}$-dependency in order to demonstrate a Laplacian.
\section{Numerical Demonstration}
\label{sec:demo}

\begin{figure}
\includegraphics[width=1.0\columnwidth]{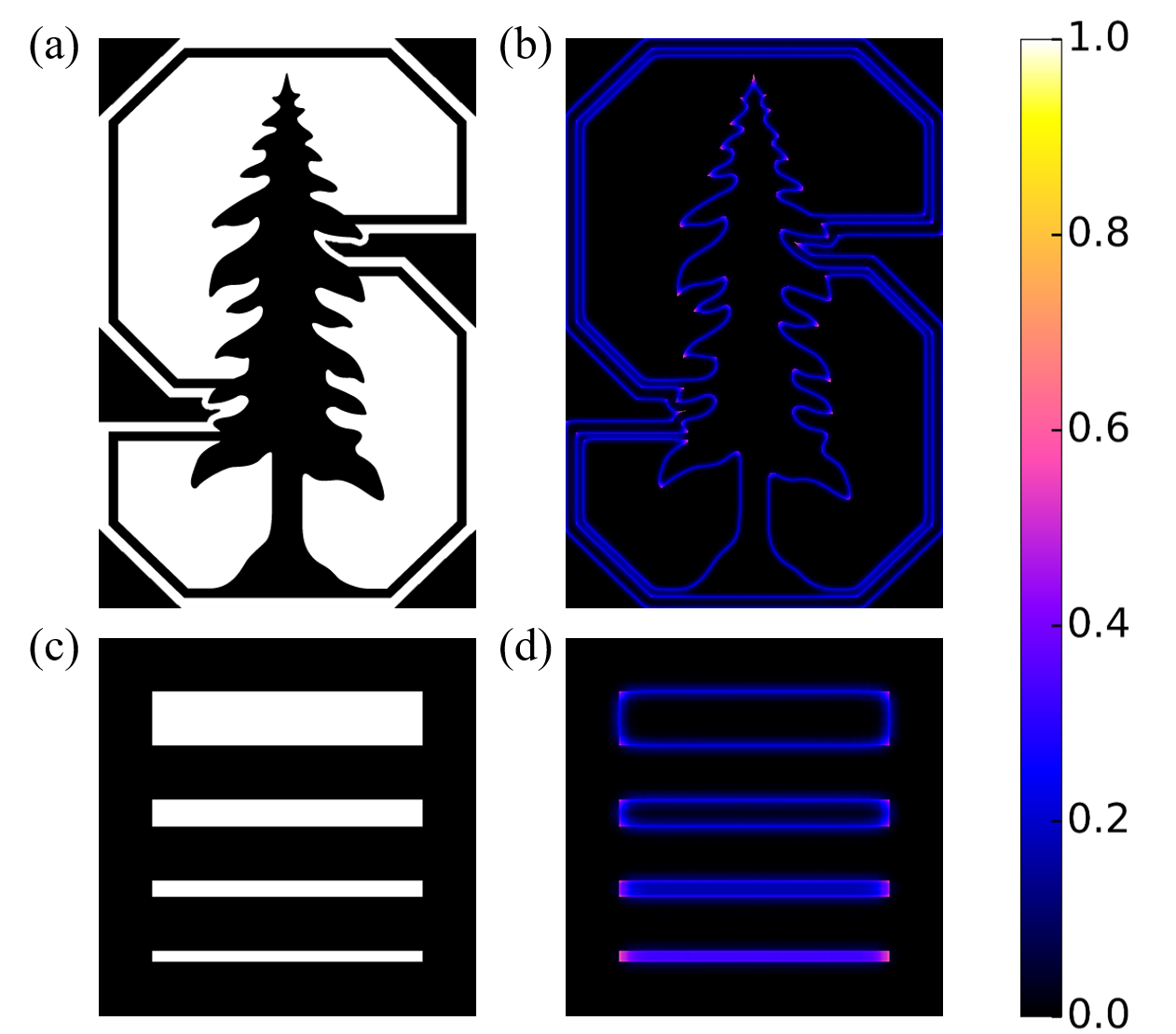}
\caption{\label{fig5}(a) Incident Stanford emblem with a size of $2610 a\times 1729 a$. (b) Calculated transmitted image, which clearly shows the edges with different orientations. (c) Incident slot patterns with length $500 a$ and width $100, 50, 30$ and $20 a$ respectively. (d) Calculated transmitted images, which show that  the spatial resolution of our design is around $30 a$.}
\end{figure}

We now show that the structure used in Figure~\ref{fig3} indeed operates as a two-dimensional Laplacian differentiator. Figure~\ref{fig5}(a) is the Stanford emblem as the incident image. Figure~\ref{fig5}(b) is the calculated transmitted image, which clearly shows all the edges with the same intensity despite different edge orientations. The ability to detect edges with different orientations is a significant advantage of an isotropic differential operator such as the two-dimensional Lapacian, as compared with the one-dimensional differential operators. The Laplacian also highlights the sharp corners, exhibiting its higher sensibility to fine details as a second-order derivative operator.\cite{Gonzales2008} 

We also test the spatial resolution of such spatial differentiator. Figure~\ref{fig5}(c) is a series of slot patterns and Figure~\ref{fig5}(d) is the calculated differentiated image. The performance of edge detection degrades as the slot width decreases. The spatial resolution of our differentiator, which is the minimum separation between the two edges that can be resolved, is around $30 a$. Considering $a=0.67\mu m$ corresponding to the resonant wavelength $\lambda_0=1.4 \mu m$, the spatial resolution is $20 \mu m$, which is sufficient for most image processing application. 
\section{Discussion and Conclusion}
In previous works, the two-dimensional Laplace operator has been implemented using holograms \cite{Guo2008,Ryle2010} and phase-shifted Bragg grating \cite{Bykov2014a}. These implementations either rely upon bulky optical components, or operate in reflection modes. In contrast, here we show this operation can be achieved with a compact optical structure in transmission mode, which is more compatible for image processing applications. 

In conclusion, we have shown that a Laplace differentiator can be implemented using  a photonic crystal slab. Such a simple photonic device
may have various applications involving image processing. Future research will be worthy to extend the Laplace differentiator to multi-frequency with multi-layer structures, enabling differentiation of color images.  

\section*{Funding Information}
This work is supported in part by Samsung Electronics, and by the U. S. Air Force Grant No.  FA9550-17-1-0002.

\section*{Acknowledgments}

The authors thank Yu Guo and Dr. Alexander Cerjan for helpful discussions.

\section*{Supplementary Materialsal Documents}

\bigskip \noindent See \href{link}{Supplementary Materials} for supporting content.

\bibliography{PCS_OSA}



\end{document}


\maketitle

\section{Derivation of the effective Hamiltonian}\label{sec:Hamiltonian}
In this section, we derive the $2\times2$ effective Hamiltonian (Equation~(10) in the primary text) near the $\Gamma$ point. We assume that the system has $C_{4v}$ symmetry. And at the $\Gamma$ point the system supports a pair of doubly degenerate states, denoted as $\ket{x}$ and $\ket{y}$, respectively. With these states as bases, the $2\times2$ Hamiltonian in the vicinity of $\Gamma$ has the following general form:
\begin{equation} 
\hat{\mathcal{H}}(\bm{k}) = \hat{\mathcal{A}}(\bm{k}) - i \hat{\mathcal{B}}(\bm{k}) 
\end{equation}
where $\hat{\mathcal{A}}(\bm{k})$ and $\hat{\mathcal{B}}(\bm{k})$ are both Hermitian and
\begin{align}
\hat{\mathcal{A}}(\bm{k}) &= f(\bm{k})\hat{\bm{\sigma}}_{+}+f^*(\bm{k})\hat{\bm{\sigma}}_{-}+g(\bm{k})\hat{\bm{\sigma}}_{z}+h(\bm{k})\hat{I}+\omega_0\hat{I} \notag \\ 
\hat{\mathcal{B}}(\bm{k}) &= r(\bm{k})\hat{\bm{\sigma}}_{+}+r^*(\bm{k})\hat{\bm{\sigma}}_{-}+s(\bm{k})\hat{\bm{\sigma}}_{z}+t(\bm{k})\hat{I}+\gamma_0\hat{I} 
\end{align}
We are interested in the lowest-order non-vanishing terms in the functions $f, g, h, r, s$ and $t$. By virtual of the two-fold degeneracy, we have
\begin{equation}
f(\bm{0})=g(\bm{0})=h(\bm{0})=r(\bm{0})=s(\bm{0})=t(\bm{0})=0
\end{equation}
Also, $g,h,s$ and $t$ are real since $\hat{\mathcal{A}}(\bm{k})$ and $\hat{\mathcal{B}}(\bm{k})$ are Hermitian.

In general, the  Hamiltonian $\hat{\mathcal{H}}(\bm{k})$ is constrained by the symmetric group $G$ of the system:
\begin{equation} \label{eq:constraint}
\forall g\in G,\qquad \hat{\mathcal{D}}(g) \hat{\mathcal{H}}(\bm{k}) \hat{\mathcal{D}}^{-1}(g) = \hat{\mathcal{H}}(\bm{D}(g)\bm{k}) 
\end{equation}
where $\hat{\mathcal{D}}(g)$ and $\bm{D}(g)$ are the representations of $g$ in the Hilbert space and the $\bm{k}$ space, respectively.

For our system, $G=C_{4v}=\{E, 2C_4, C_2, 2\sigma_v, 2\sigma_d\}.$ As we choose $\ket{x}$ and $\ket{y}$ as bases for the Hilbert space and $\bm{k_x}$ and $\bm{k_y}$ for the $\bm{k}$ space, the representations $\hat{\mathcal{D}}(g)$ and $\bm{D}(g)$ have the same matrix forms. Now, we consider the constraint on $\hat{\mathcal{H}}(\bm{k})$ from each symmetry element in $G$.
\begin{enumerate}
\item {\bf Inversion symmetry $C_2$.} 
\begin{equation}
\hat{\mathcal{D}}(C_2) = \bm{D}(C_2) \equiv \bm{\Pi}= \begin{pmatrix}
-1 & 0 \\
0 & -1
\end{pmatrix}
\end{equation}
From Equation~(\ref{eq:constraint}) we have:
\begin{align}
&f(\bm{k})=f(-\bm{k}),\quad g(\bm{k})=g(-\bm{k}),\quad h(\bm{k})=h(-\bm{k}) \notag \\ 
&r(\bm{k})=r(-\bm{k}),\quad s(\bm{k})=s(-\bm{k}),\quad t(\bm{k})=t(-\bm{k}) 
\end{align}
Thus, the lowest order terms in $f, g, h, r, s$ and $t$ must all be quadratic.
\item {\bf Four-fold rotational symmetry $C_4$.}
\begin{equation}
\hat{\mathcal{D}}(C_4) = \bm{D}(C_4) \equiv \bm{R} = \begin{pmatrix}
0 & 1 \\
-1 & 0
\end{pmatrix}
\end{equation}
From Equation~(\ref{eq:constraint}) we have:
\begin{align}
&f^*(\bm{k})=-f(\bm{R}\bm{k}),\quad g(\bm{k})=-g(\bm{R}\bm{k}),\quad h(\bm{k})=h(\bm{R}\bm{k}) \notag \\ 
&r^*(\bm{k})=-r(\bm{R}\bm{k}),\quad s(\bm{k})=-s(\bm{R}\bm{k}),\quad t(\bm{k})=t(\bm{R}\bm{k}) 
\end{align}
\item {\bf Mirror symmetry $\sigma_v$.}
\begin{equation}
\hat{\mathcal{D}}(\sigma_v) = \bm{D}(\sigma_v) \equiv \bm{M_y} = \begin{pmatrix}
1 & 0 \\
0 & -1
\end{pmatrix}
\end{equation}
Here we consider the mirror plane perpendicular to the y-axis. From Equation~(\ref{eq:constraint}) we have:
\begin{align}
&f(\bm{k})=-f(\bm{M_y}\bm{k}),~ g(\bm{k})=g(\bm{M_y}\bm{k}),~ h(\bm{k})=h(\bm{M_y}\bm{k}) \notag \\ 
&r(\bm{k})=-r(\bm{M_y}\bm{k}),~ s(\bm{k})=s(\bm{M_y}\bm{k}),~ t(\bm{k})=t(\bm{M_y}\bm{k})~ 
\end{align}
\end{enumerate}

The other mirror symmetry $\sigma_d$ doesn't provide new constraints as it can be realized with a combination of $\sigma_v$ and  $C_{4v}$. 

Combining all the symmetry requirements, we determine the forms of the lowest order terms in all the functions. As an example, we consider the function $f(\bm{k})$, which can be expanded as
\begin{equation}
f(\bm{k}) = C_x k_x^2 + C_y k_y^2 + C_{xy} k_x k_y
\end{equation}
where all the $C$ coefficients are in general complex. Since $f(\bm{k})=-f^*(\bm{R}\bm{k})=-f(\bm{M_y}\bm{k})$, we have $C_x = C_y = 0$, $C_{xy}=C^*_{xy}\equiv C$. Thus $f(\bm{k})=C k_x k_y$, where $C$ is real. Repeating the procedures for all the functions, we summarize the results below: 
\begin{align}
&f(\bm{k}) = C k_x k_y,\qquad &r(\bm{k})= C' k_x k_y \notag\\
&g(\bm{k}) = B (k_x^2-k_y^2),\qquad &s(\bm{k}) = B' (k_x^2-k_y^2) \notag\\
&h(\bm{k}) = A (k_x^2+k_y^2),\qquad &t(\bm{k}) = A' (k_x^2+k_y^2)
\end{align}
where all the coefficients are real.

Therefore, we obtain the Hamiltonian in Equation~(10) of the main text.:
\begin{equation} \label{eq:H-k}
\hat{\mathcal{H}}(\bm{k}) = (\omega_0 - i \gamma_0 + a |\bm{k}|^2)\hat{\mathbf{I}} + b (k_x^2-k_y^2) \hat{\bm{\sigma}}_z+ c k_x k_y \hat{\bm{\sigma}}_x
\end{equation}
where $a=A-iA',~b=B-iB',~c=C-iC'$ are the complex coefficients. 

\section{Validation of the effective Hamiltonian}\label{sec:Validation_Hamiltonian}
\begin{figure}
\includegraphics[width=0.9\columnwidth]{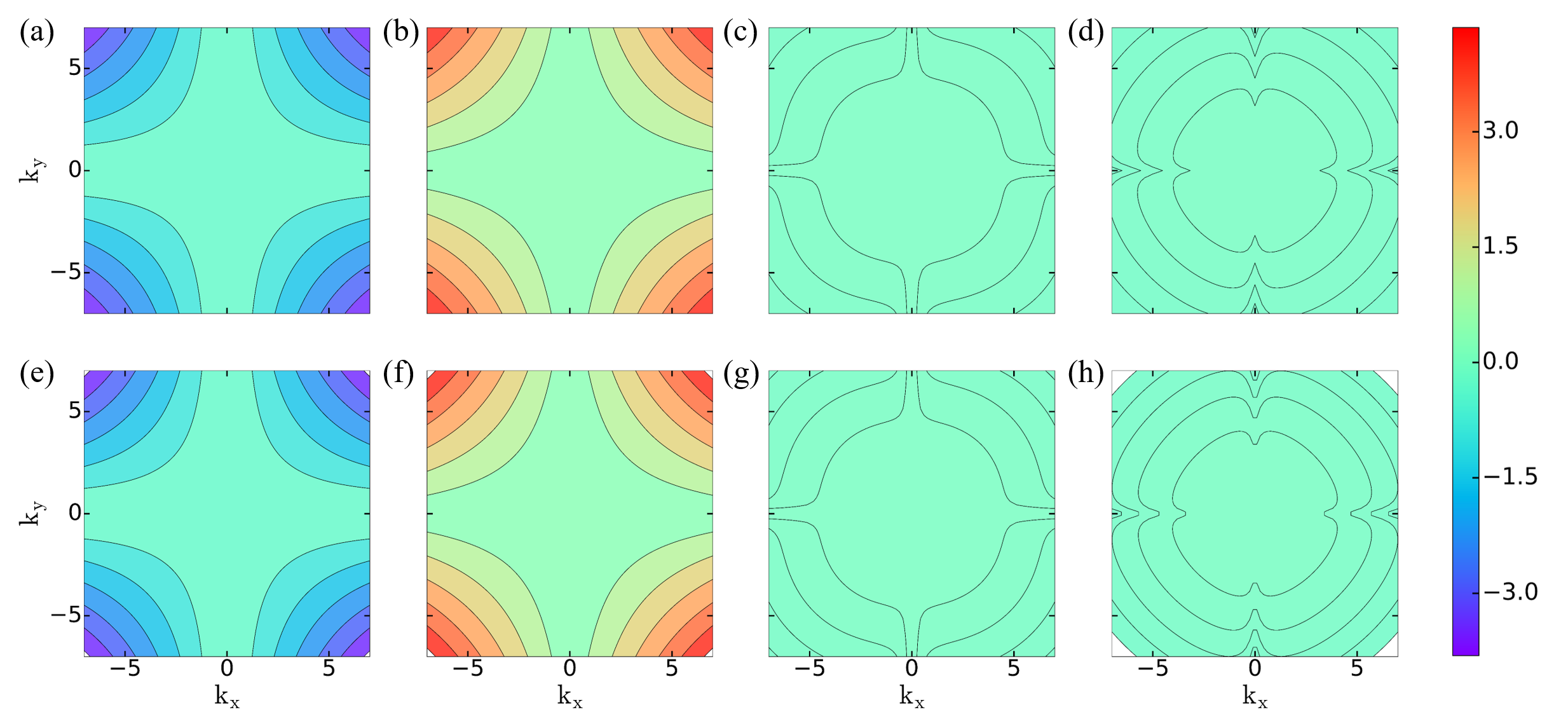}
\centering
\caption{\label{fig:SI_model_anisotropy}Band structure of the photonic crystal slab near frequency $\omega_0=0.38749 \times 2\pi c/a$. The structure has the dielectric constant  $\epsilon=12$, the thickness $d=0.55a$, and the radius $r=0.111a$ of the holes. (a-d) Contour plots of the band structure numerically calculated using guided-mode expansion method. (a) $(\omega-\omega_0)\textrm{-}(k_x,k_y)$ for the lower band. (b) $(\omega-\omega_0)\textrm{-}(k_x,k_y)$ for the upper band. (c) $(\gamma-\gamma_0)\textrm{-}(k_x,k_y)$ for the lower band. (d) $(\gamma-\gamma_0)\textrm{-}(k_x,k_y)$ for the upper band. $\gamma_0= 2.0\times10^{-4}\times 2\pi c/a$. The $(\omega-\omega_0)$ and $(\gamma-\gamma_0)$ contours for both bands are anisotropic. (e-h) Corresponding contour plots of the band structure analytically calculated from the effective Hamiltonian (Equation (\ref{eq:H-k})). The complex coefficients $a=0.33
-0.16i,b=0.24-0.16i,c=8.78
-0.08i$ are fitted from the band dispersions along $\Gamma\textrm{-}X$ and $\Gamma\textrm{-}M$ direction. The numerical results agree with the analytical expressions excellently. In all the plots $k_x$ and $k_y$ are in the units of $10^{-3}\times2\pi/a$, $(\omega-\omega_0)$ and $(\gamma-\gamma_0)$ are in the units of $10^{-4}\times2\pi c/a$.
}
\end{figure}

\begin{figure}
\includegraphics[width=0.9\columnwidth]{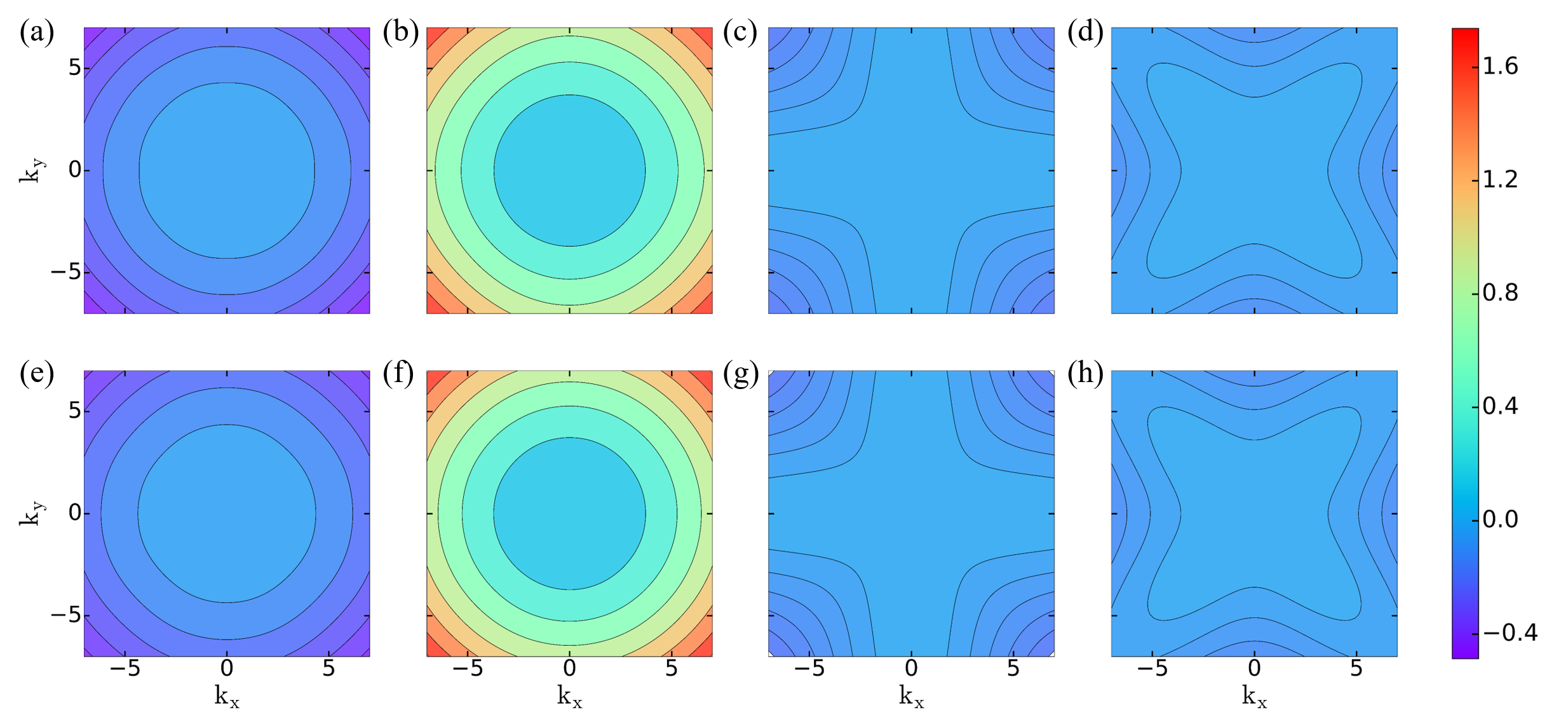}
\centering
\caption{\label{fig:SI_model}Band structure of the photonic crystal slab near frequency $\omega_0=0.47656 \times 2\pi c/a$. The structure has the dielectric constant  $\epsilon=12$, the thickness $d=0.55a$, and the radius $r=0.111a$ of the holes. (a-d) Contour plots of the band structure numerically calculated using guided-mode expansion method. (a) $(\omega-\omega_0)\textrm{-}(k_x,k_y)$ for the lower band. (b) $(\omega-\omega_0)\textrm{-}(k_x,k_y)$ for the upper band. (c) $(\gamma-\gamma_0)\textrm{-}(k_x,k_y)$ for the lower band. (d) $(\gamma-\gamma_0)\textrm{-}(k_x,k_y)$ for the upper band. $\gamma_0= 1.3\times10^{-4}\times 2\pi c/a$. The $(\omega-\omega_0)$ contours for both bands are almost completely circular, while the $(\gamma-\gamma_0)$ contours are anisotropic. Nonetheless, $(\gamma-\gamma_0)$ are much smaller than $(\omega-\omega_0)$ and thus don't affect the isotropy of the band structure much. (e-h) Corresponding contour plots of the band structure analytically calculated from the effective Hamiltonian (Equation (\ref{eq:H-k})). The complex coefficients $a=0.68-0.14i,b=1.11-0.12i,c=2.24+0.15i$ are fitted from the band dispersions along $\Gamma\textrm{-}X$ and $\Gamma\textrm{-}M$ direction. The numerical results agree with the analytical expressions excellently. In all the plots $k_x$ and $k_y$ are in the units of $10^{-3}\times2\pi/a$, $(\omega-\omega_0)$ and $(\gamma-\gamma_0)$ are in the units of $10^{-4}\times2\pi c/a$.
}
\end{figure}

In this section we provide quantitative validation of the effective Hamiltonian. Figure (\ref{fig:SI_model_anisotropy}) plots the band structure of the photonic crystal slab near frequency $\omega_0=0.38749 \times 2\pi c/a$. Figure (\ref{fig:SI_model_anisotropy}) (a-d) exhibit contour plots of $(\omega-\omega_0)$ and $(\gamma-\gamma_0)$ for the two bands numerically calculated using guided-mode expansion method, while (e-h) show the corresponding analytical results from the effective Hamiltonian (Equation (\ref{eq:H-k})). The complex coefficients $a=0.33-0.16i,b=0.24-0.16i,c=8.78-0.08i$ are fitted from the band dispersions along $\Gamma\textrm{-}X$ and $\Gamma\textrm{-}M$ direction. Figure (\ref{fig:SI_model}) plots the numerical and analytical results of the band structure of the same photonic crystal slab near frequency $\omega_0=0.47656 \times 2\pi c/a$.  The fitted complex coefficients are $a=0.68-0.14i,b=1.11-0.12i,c=2.24+0.15i$. In both cases, the numerical results agree with the analytical expressions excellently, which provide the validations of the effective Hamiltonian.

\section{Transmittance of arbitrary polarized states}\label{sec:DensityMatrix}
In this section we compute the transmission of an arbitrarily polarized state,  given the transmission response of the system for $S$ and $P$ polarized light.

The transmission response of the system is described by the $S$ matrix:
\begin{equation}
S = \begin{pmatrix}
t_{ss} & t_{sp} \\
t_{ps} & t_{pp}
\end{pmatrix}
\end{equation}
where $t_{ss}, t_{sp}, t_{ps}$ and $t_{pp}$ are complex. Here we adopt $\ket{s}$ and $\ket{p}$ as the basis states. For a general input state $\ket{e}$, the output state is $S\ket{e}$.

Consider an input beam with its polarization described by a  polarization density matrix $\rho$, normalized such that $I = \operatorname{tr} \rho$ is the incident light intensity. The transmitted state is then described by $\rho' = S \rho S^\dagger$, 
with the transmission defined as: 
\begin{equation} \label{eq:T}
T = \operatorname{tr} \rho' / \operatorname{tr} \rho 
\end{equation}

Now let's consider some specific  examples of incident states:
\begin{enumerate}
\item {\bf S polarized light} 
\begin{align}
\rho &= I \begin{pmatrix}
1 & 0 \\
0 & 0
\end{pmatrix}\notag \\
\rho' &= I \begin{pmatrix}
|t_{ss}|^2 & t_{ss} t_{ps}^* \\
t_{ss}^* t_{ps} & |t_{ps}|^2
\end{pmatrix}\notag\\
T &\equiv |t_s|^2 = |t_{ss}|^2+|t_{ps}|^2 \label{eq:ts}
\end{align}
Similarly, for P polarized light,
\begin{equation}
|t_p|^2 = |t_{sp}|^2+|t_{pp}|^2 \label{eq:tp}
\end{equation}

\item {\bf Unpolarized light} 
\begin{align}
\rho &= \frac{I}{2} \begin{pmatrix}
1 & 0 \\
0 & 1
\end{pmatrix}\notag\\
\rho' &= \frac{I}{2}  \begin{pmatrix}
|t_{ss}|^2 + |t_{sp}|^2 & t_{ss} t_{ps}^* + t_{sp}t_{pp}^* \\
t_{ss}^* t_{ps}+ t_{sp}^*t_{pp} & |t_{ps}|^2 + |t_{pp}|^2
\end{pmatrix}\notag\\
T &\equiv |t_u|^2 = \frac{1}{2}(|t_{ss}|^2+|t_{ps}|^2+|t_ps|^2+|t_{pp}|^2) \label{eq:tu}
\end{align}

From Equation~(\ref{eq:ts}-\ref{eq:tu}) we have:
\begin{equation}
|t_u| = \sqrt{\frac{|t_s|^2+|t_p|^2}{2}} 
\end{equation}
which is Equation~(18) in the main text.

\item {\bf Left Circularly polarized light} 
\begin{align}
\rho &= \frac{I}{2} \begin{pmatrix}
1 & -i \\
i & 1
\end{pmatrix}\notag\\
\rho' &= \frac{I}{2}  \begin{pmatrix}
|t_{ss}|^2 + |t_{sp}|^2 + i(t_{sp} t_{ss}^* - t_{sp}^* t_{ss}) & t_{ss} t_{ps}^* + t_{sp}t_{pp}^* + i(t_{sp} t_{ps}^*-t_{ss}t_{pp}^*) \\
t_{ss}^* t_{ps}+ t_{sp}^*t_{pp} - i(t_{sp}^* t_{ps}-t_{ss}^*t_{pp}) & |t_{ps}|^2 + |t_{pp}|^2 +i(t_{ps}^* t_{pp} - t_{ps} t_{pp}^*)
\end{pmatrix}\notag\\
T & \equiv |t_{l}|^2 = \frac{1}{2}[|t_{ss}|^2+|t_{ps}|^2+|t_ps|^2+|t_{pp}|^2 + i (t_{sp}t_{ss}^*-t_{sp}^*t_{ss}+t_{ps}^*t_{pp}-t_{ps}t_{pp}^*)]\label{eq:tl}
\end{align}
which differs from Equation~(\ref{eq:tu}) only in the interference term.
\end{enumerate}

Note that for our system with an isotropic band structure, $S$ is diagonal:
\begin{equation}
S = \begin{pmatrix}
t_{ss} & 0 \\
0 & t_{pp}
\end{pmatrix}
\end{equation}
Then the interference term in Equation~(\ref{eq:tl}) disappears and we have $|t_l|=|t_u|$. So the differentiation performance will be the same under illumination of unpolarized or circularly polarized light. 

\section{Transmittance at larger wavevectors}\label{sec:t_k}
\begin{figure}
\includegraphics[width=0.9\columnwidth]{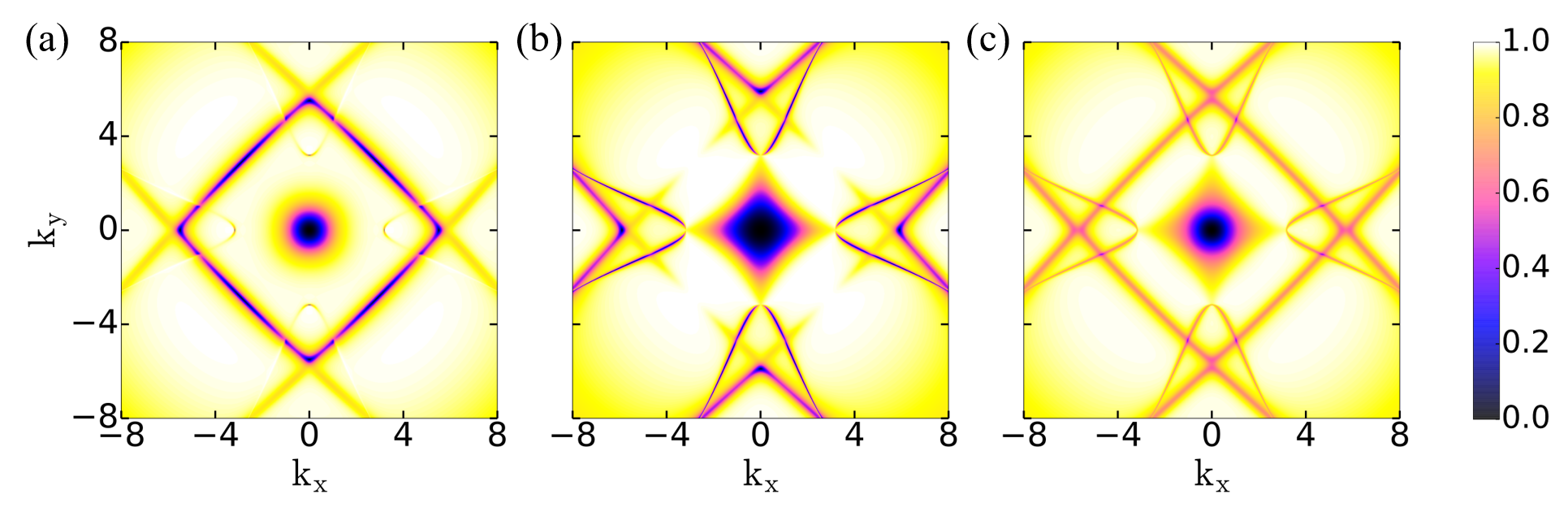}
\centering
\caption{\label{fig:t_map}Transmittance $|t|$ as a function of $k_x$ and $k_y$ at frequency $\omega_0 = 0.47656 \times 2\pi c/a$ for (a) $S$, (b) $P$ and (c) unpolarized light, which are all isotropic near $\Gamma$, which is the major relevant wavevector region for image processing, but anisotropic at larger $\bm{k}$. $k_x$ and $k_y$ are in the units of $10^{-2}\times2\pi/a$.}
\end{figure}
Figure~(\ref{fig:t_map}) plots the transmittance $|t|$ as a function of $k_x$ and $k_y$ for $S$, $P$ and unpolarized light in a larger wavevector region as compared to Figure 4 of the primary text. All $|t|$ are isotropic near the $\Gamma$ point. They become anisotropic at larger wavevector $\bm{k}$. The background transmittances are all $1$. The occurrence of transmittance dips at larger $\bm{k}$ are due to other guided resonance modes.   

We calculate the performance of our differentiator based on the transmittance shown here, which covers all the relevant wavevector regions for the incident images that we assumed. 
